\documentclass[cits]{PoS}
\usepackage{fontenc}
\usepackage{graphicx}
\usepackage{times}
\usepackage{mathptmx}
\usepackage{amssymb}

\title{The Central 3 kpc of NGC 5850}

\ShortTitle{The Central 3 kpc of NGC 5850}

\author{\speaker{M. Bremer}$^a$\thanks{This work is carried out within the Collaborative Research Council 956, sub-project [A2], funded by the Deutsche Forschungsgemeinschaft (DFG). M.B., J.S, J.Z., and S.F. are grateful for the travel support granted by the Group of 8 (Go8) of Australia and the German Academic Exchange Service (DAAD). This research has made use of the NASA/IPAC Extragalactic Database (NED) which is operated by the Jet Propulsion Laboratory, California Institute of Technology, under contract with the National Aeronautics and Space Administration.}, J. Scharw\"achter$^b$, A. Eckart$^{a,c}$, J. Zuther$^a$, S. Fischer$^a$, M. Valencias-S.$^a$, F. Combes$^b$, and S. Garcia-Burillo$^d$\\
\llap{$^a$} I. Institute of Physics, University of Cologne\\
Z\"ulpicher Strasse 77, 50937 Cologne, Germany\\
\llap{$^b$} Observatoire de Paris, LERMA (CNRS: UMR 8112)\\
61 Av. de l'Observatoire, 75014, Paris, France\\
\llap{$^c$} Max-Planck-Institut f\"ur Radioastronomie\\
Auf dem H\"ugel 69, 53121 Bonn, Germany\\
\llap{$^d$} Observatorio Astronomico Nacional (OAN), Observatorio de Madrid\\
Alfonso XII, 3, Madrid, Spain\\
        E-mail: \email{mbremer@ph1.uni-koeln.de}}

\abstract{NGC 5850 is a nearby $\mathrm{\left(z=0.0085\right)}$ early type spiral galaxy classified as LINER. It is considered as a prototype double-barred system. Our optical Integral Field Spectroscopic (IFS) data of the central $21 \times 19~\mathrm{arcsec^{2}}$ of NGC 5850 show extended LINER-like emission which we ascribe to the presence of a hot and evolved stellar population, possibly together with a faint AGN. Additionally NGC 5850 shows extended `composite' ionization patterns, likely to stem from a mixture of LINER-like ionization and photoionization by star formation. The kinematics of the gas deviates strongly from a simple rotational structure.}

\FullConference{Nuclei of Seyfert galaxies and QSOs - Central engine \& conditions of star formation,\\
		November 6-8, 2012\\
		Max-Planck-Insitut f\"ur Radioastronomie (MPIfR), Bonn,  			Germany}

\begin{document}

\section{Introduction}
\label{intro}
Low Ionization Nuclear Emission Region (LINER) \cite{Heckman1980} galaxies have caused controversy for decades. The low luminosity of these objects could principally come from compact stellar populations in their centers \cite{Pogge}. Since active galactic nuclei (AGN) are commonly objects of high luminosity with strong ionization features, this raises the question whether LINERs represent the low luminosity part of the family of active galactic nuclei (LLAGN) or are not AGN at all.\
Not only the low luminosity poses a challenge. The typical emission line ratios of LINERs do not allow an unambiguous identification of the ionization mechanism without any additional information. The line ratios can be produced as well by shocks \cite{Dopita_shocks}, cooling flows \cite{Voit} or post-AGB stars \cite{Terlevich85}. Unresolved point sources can produce extended emission as well \cite{Yan} making it even more complicated to distinguish between photoionization by an AGN or by other mechanisms. Consequently, the role of LINERs in the context of AGN evolution is not clear yet.\

Here, we present the results of the analysis of the VIMOS IFU observations of the central $\sim3~\mathrm{kpc}$ of the LINER galaxy NGC 5850. We examine present ionization mechanisms and look for signs of AGN feeding and feedback. NGC 5850 is a nearby $\mathrm{\left(z=0.0085\right)}$ prototype double-barred early-type spiral galaxy (e.g. \cite{Buta}), with a position angle of $\approx 140~\mathrm{degrees}$ \cite{deVaucouleurs} and inclined by $\mathrm{i \approx 37~degrees}$. The large scale (primary) and the inner (secondary) bar differ in their P.A. by $-67$ degrees, with  $\mathrm{P.A._{primary} \approx 116.3~degrees}$ and $\mathrm{P.A._{secondary} \approx 49.3~degrees}$ \cite{Lourenso}. Instead of a secondary bar, \cite{Moiseev} proposed the presence of gaseous polar disk. They also found the core region to be kinematically decoupled (see also \cite{Leon}). \cite{Higdon} report the possibility of a high velocity encounter with NGC 5846 $<200~\mathrm{Myr}$ ago. The X-ray luminosity is low $\left(\mathrm{L_X\left(0.5-3~keV\right) = 10^{40.36}~erg~s^{-1}}\right)$ indicating a low activity level of the AGN. Therefore, \cite{Fabbiano} did not consider it likely to be AGN dominated. Radio observations were performed by \cite{Leon}, \cite{Hummel} and \cite{Condon}. The latter conclude on a star formation dominated core.\

Because of its low luminosity and its proximity NGC 5850 is a good candidate to examine the mechanisms of AGN-feeding/AGN feedback. As such it is part of the NUGA (NUclei of GAlaxies) sample, a $^{12}\mathrm{CO}$ survey focussing on the core areas of LLAGN (e.g. \cite{NUGA_Garcia_Burillo, NUGA_Combes}).\

\section{Morphology}
\label{morph}

\begin{figure}
  \centering
  \includegraphics[width=0.32\textwidth]{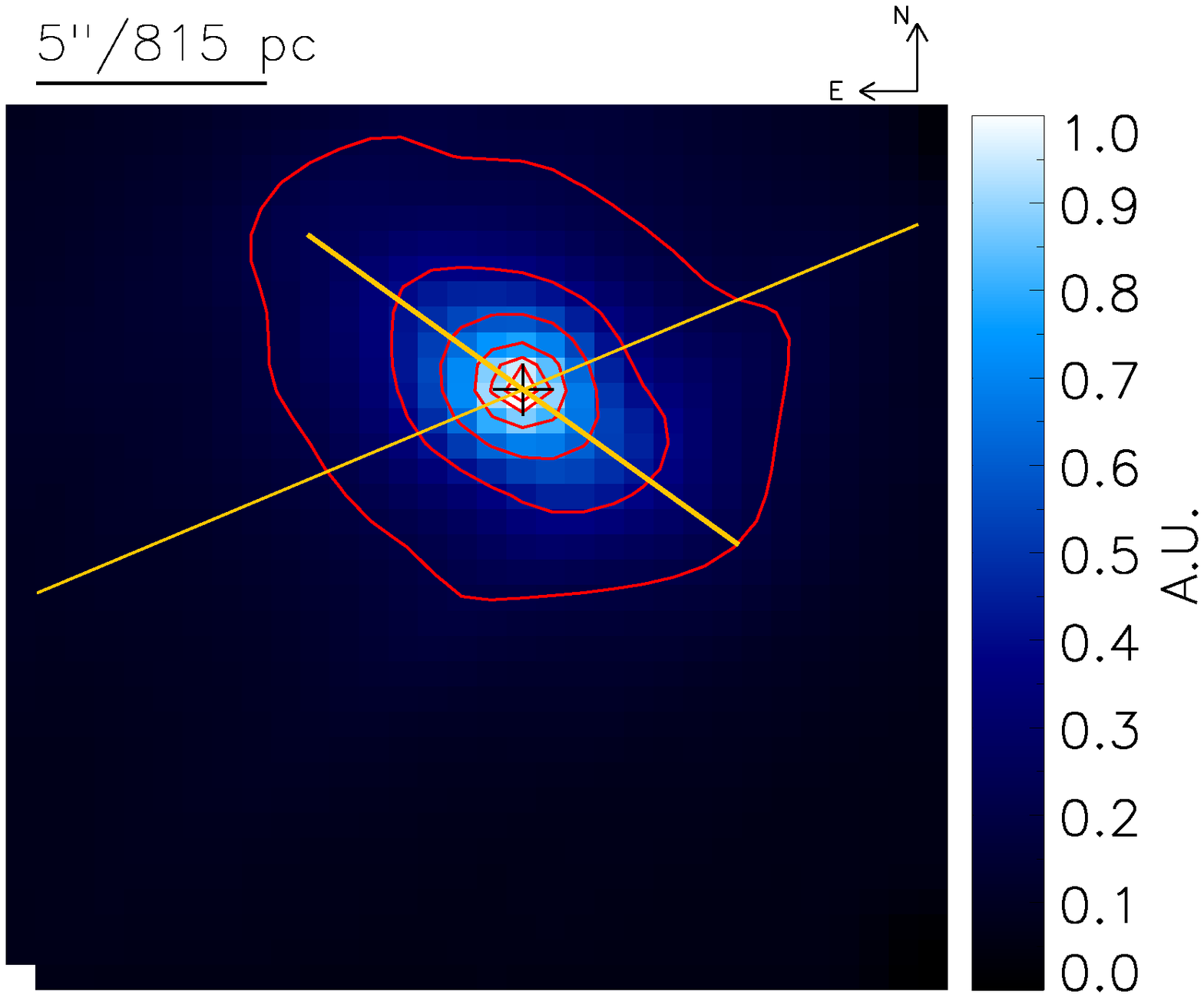}
  \includegraphics[width=0.32\textwidth]{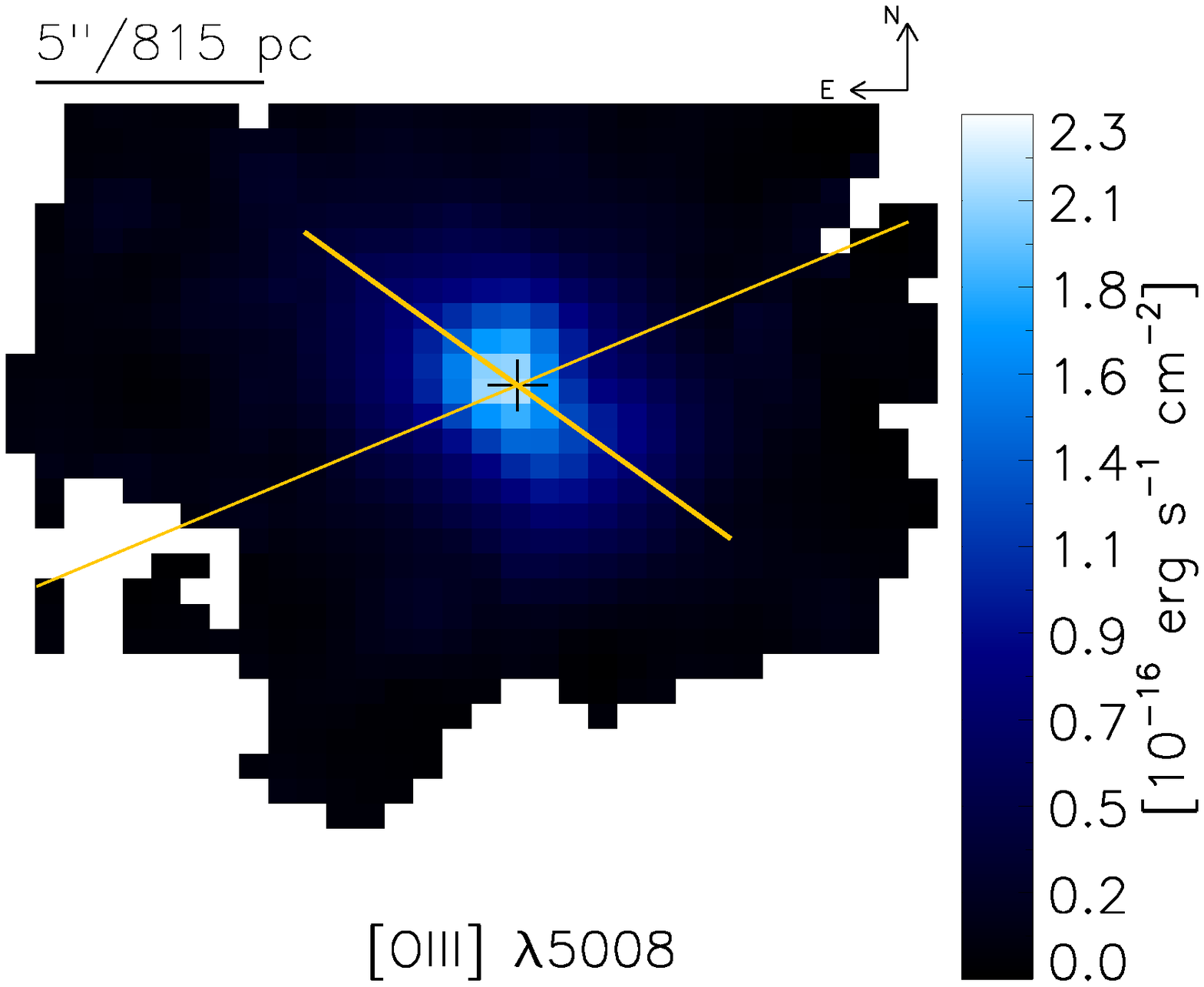}
  \includegraphics[width=0.32\textwidth]{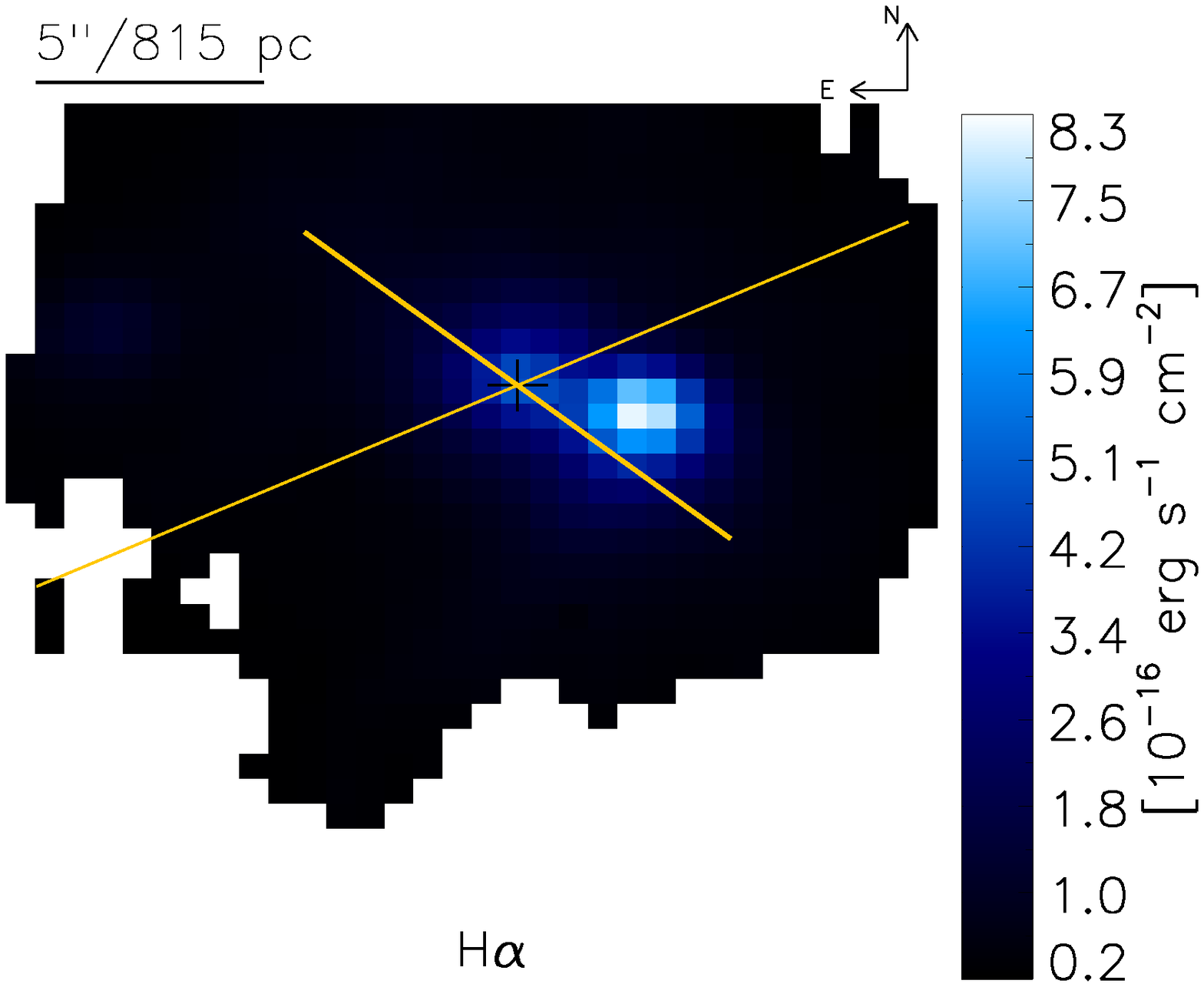}
  \caption{\textbf{Left:} Continuum image ($6000$ to $6100$~\AA) of NGC 5850. The long line represents the major axis of the primary bar, the short line represents the secondary bar. The isointensity contours are at levels of 20, 40, 60, 80, 90 and 95 \% of maximum value). \textbf{Center:} The [OIII] $\lambda 5008$ emission line intensity map. \textbf{Right:} H$\alpha$ emission line intensity map.}
 \label{fig:emission}
 \end{figure}

In Fig. \ref{fig:emission} we present several emission line maps. The continuum image derived from our observations is elongated along the major axis of the secondary bar whose P.A. is given by $\mathrm{49.3~degrees}$ by \cite{Lourenso} in agreement with our own elliptical fits. The continuum intensity peaks in the center. The [OIII] $\lambda 5008$ emission line intensity shows similar structures. In contrast, the H$\alpha$ emission has a clear secondary peak located off-core at a distance of $\sim 345~\mathrm{pc}$, indicating an HII region. A third very weak H$\alpha$ peak can be found far east ($\approx 1627~\mathrm{pc}$) of the center. The H$\beta$ emission line map confirms these structures. The emission line maps of [NII] $\lambda 6585$ and [SII] $\lambda\lambda 6718,~6732$ (not shown here) are similar to [OIII] but expose a small additional contribution at the location of the central HII-region.

\section{Kinematics}
\label{kinematics}

\begin{figure}
  \centering
  \includegraphics[width=0.32\textwidth]{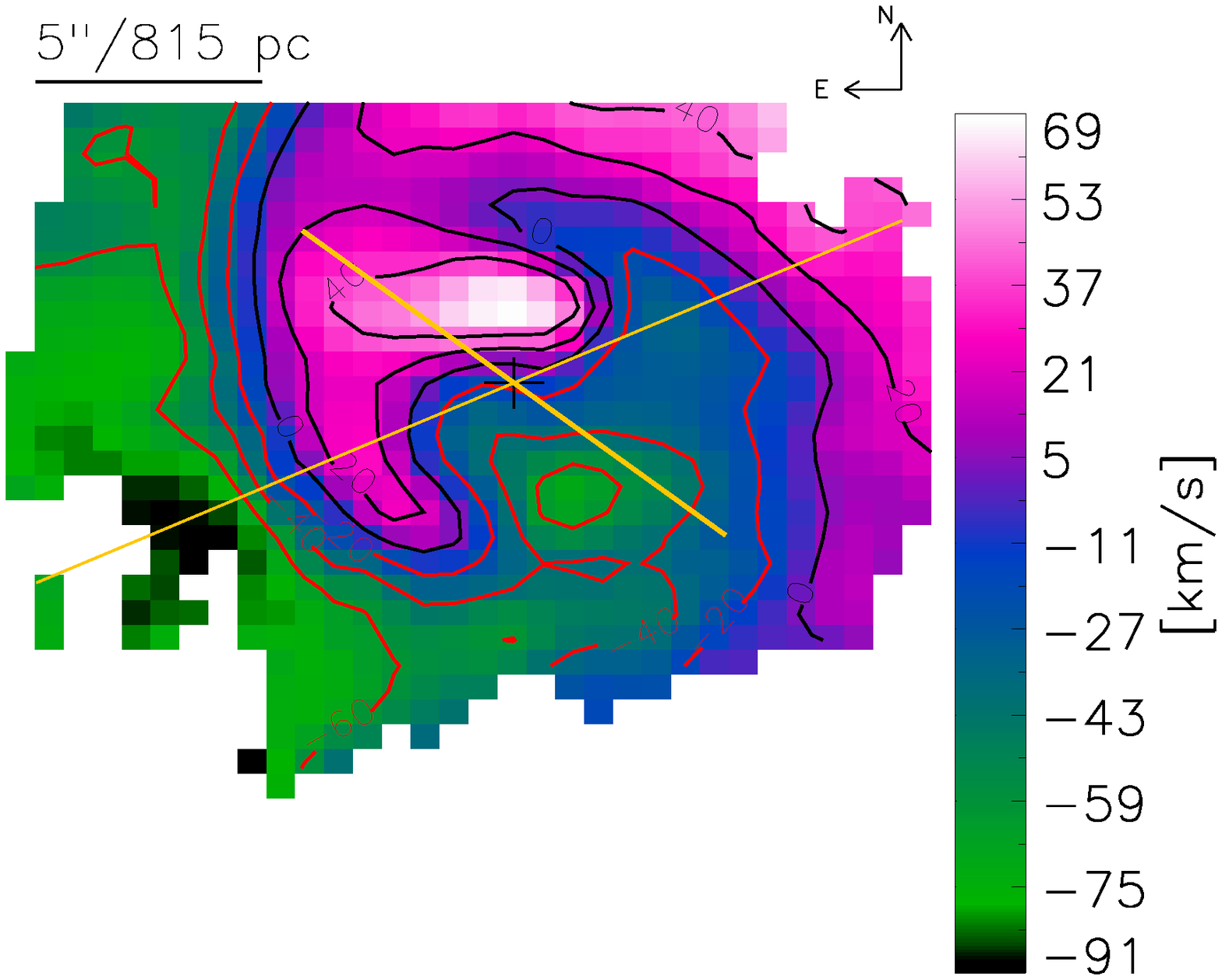}
  \includegraphics[width=0.32\textwidth]{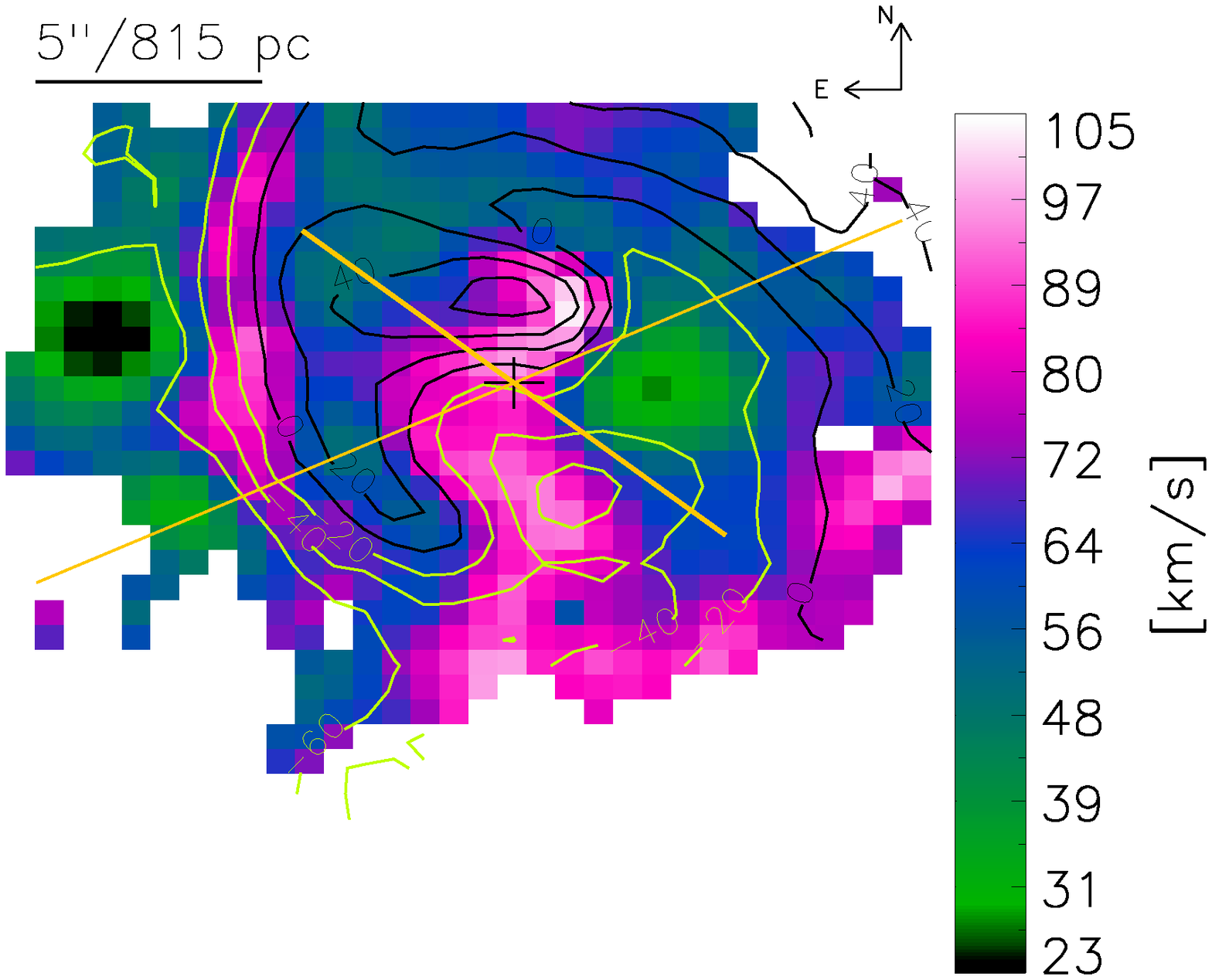}
  \caption{\textbf{Left:} H$\alpha$ LOSV map with isovelocity contours at $-60$, $-40$, $-20$, $0$, $20$, and $40~\mathrm{km~s^{-1}}$. \textbf{Right:} The H$\alpha$ LOSV dispersion map with isovelocity contours as in left panel.}
 \label{fig:kinematics}
 \end{figure}

The H$\alpha$ line of sight velocity (LOSV) map in the left panel of Fig. \ref{fig:kinematics} is characterized by a strong spiral-like structure. The positive and negative velocity peaks can be found north-east and south-west of the center, respectively. The kinematic influence of the secondary bar is indicated by the typical s-shape of the isovelocity contours in the center. Furthermore, the distinctiveness of the kinematics within the radius of the secondary bar compared to more distant regions implies the core to be largely kinematically decoupled.\

The H$\alpha$ LOSV dispersion map (right panel of Fig. \ref{fig:kinematics}) is relatively flat. However, on one hand it exposes low values at the locations of HII-regions as expected for star-forming regions. On the other hand the high values follow the steep velocity gradients. This is likely due to beam smearing, an effect caused by insufficient spatial resolution. Nevertheless, a contribution from slow shocks cannot be ruled out.\

\begin{figure}
  \centering
  \includegraphics[width=0.32\textwidth]{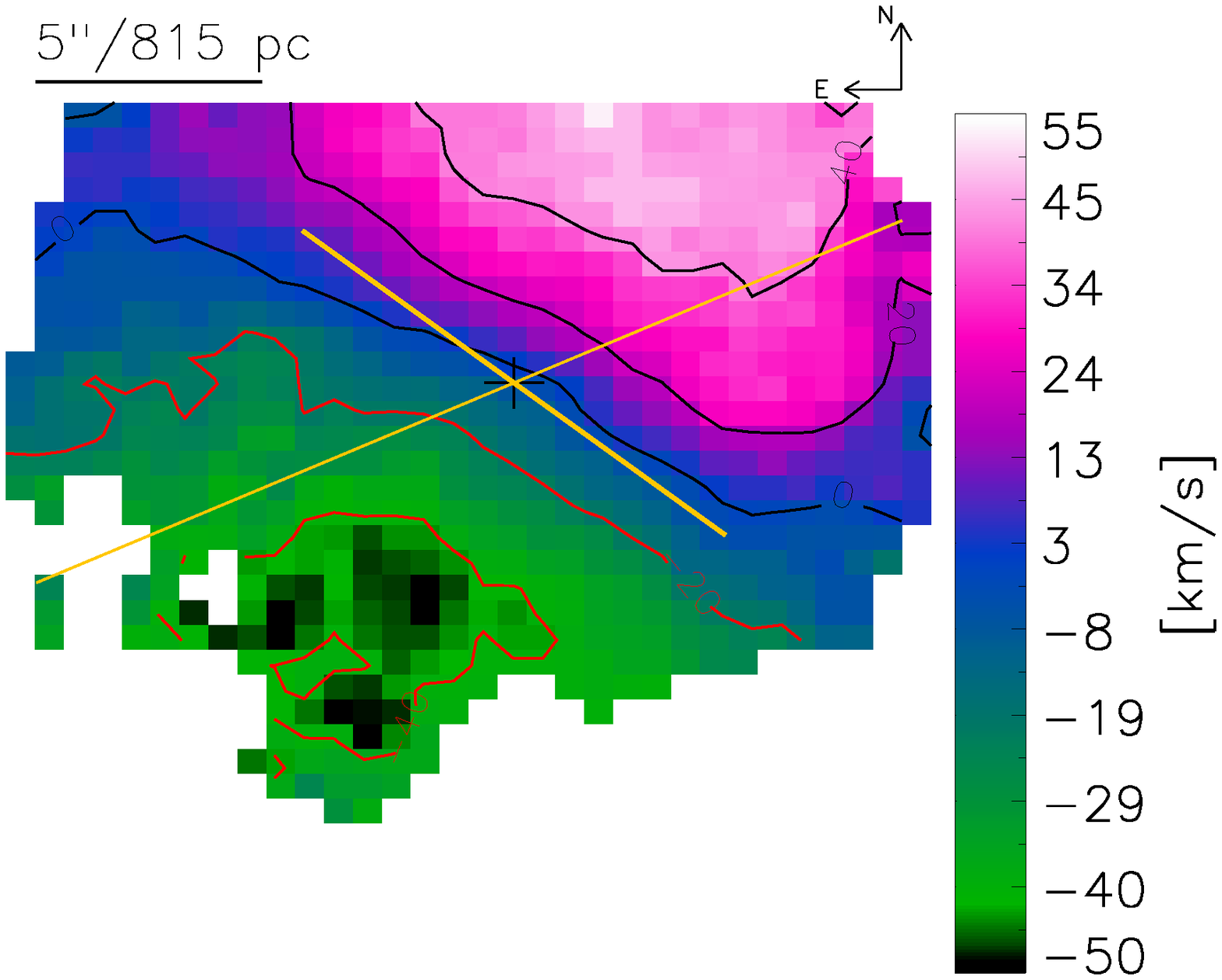}
  \includegraphics[width=0.32\textwidth]{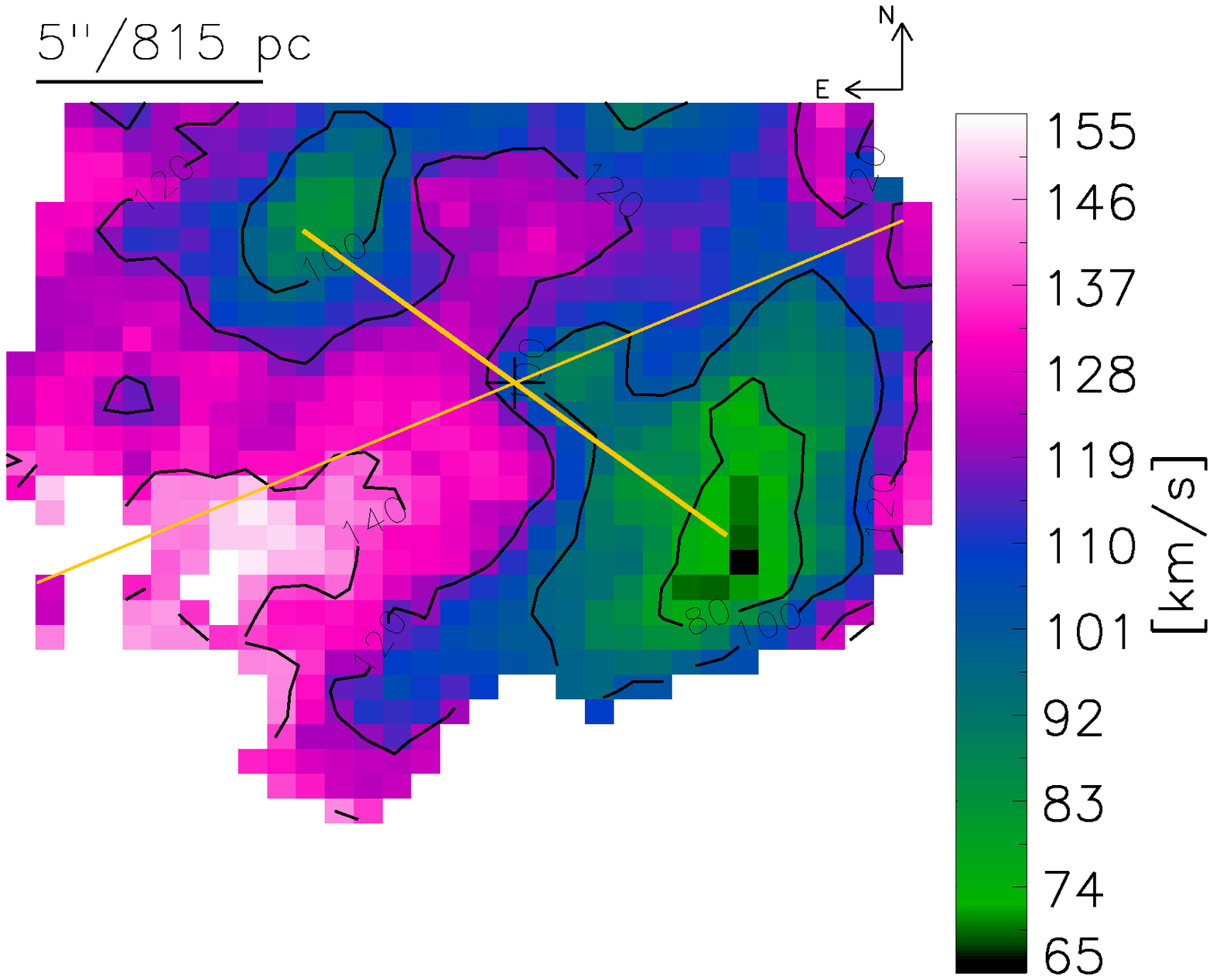}
  \caption{\textbf{Left:} The stellar LOSV map as derived by the stellar population synthesis code STARLIGHT. \textbf{Right:} The stellar LOSV dispersion map.}
 \label{fig:stellar_kin}
 \end{figure}

In the left panel of Fig. \ref{fig:stellar_kin} we present the stellar LOSV map as derived by the stellar population synthesis code STARLIGHT \cite{starlight}. It is characterized by almost only rotational properties with hardly any influence of the secondary bar, implying the inner bar to be mostly gaseous. The stellar LOSV dispersion map confirms the detection of the so-called `$\sigma$-hollows' \cite{sigma_hollows}, regions close to the tips of the secondary bar with low stellar LOSV dispersion values. They ascribe this to a contrast effect between bulge and bar.

\section{Excitation}
\label{excitation}

\begin{figure}
  \centering
  \includegraphics[width=0.32\textwidth]{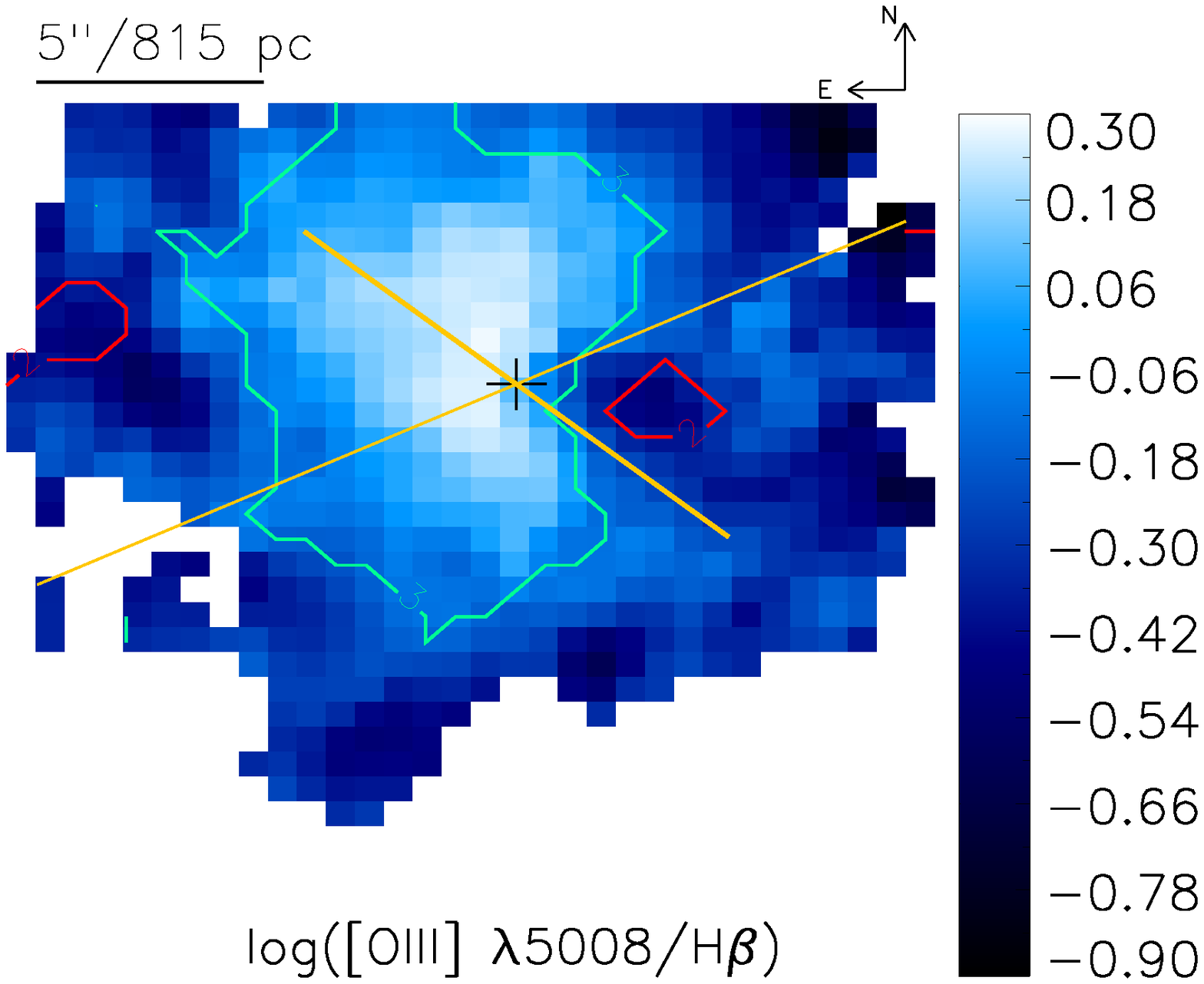}
  \includegraphics[width=0.32\textwidth]{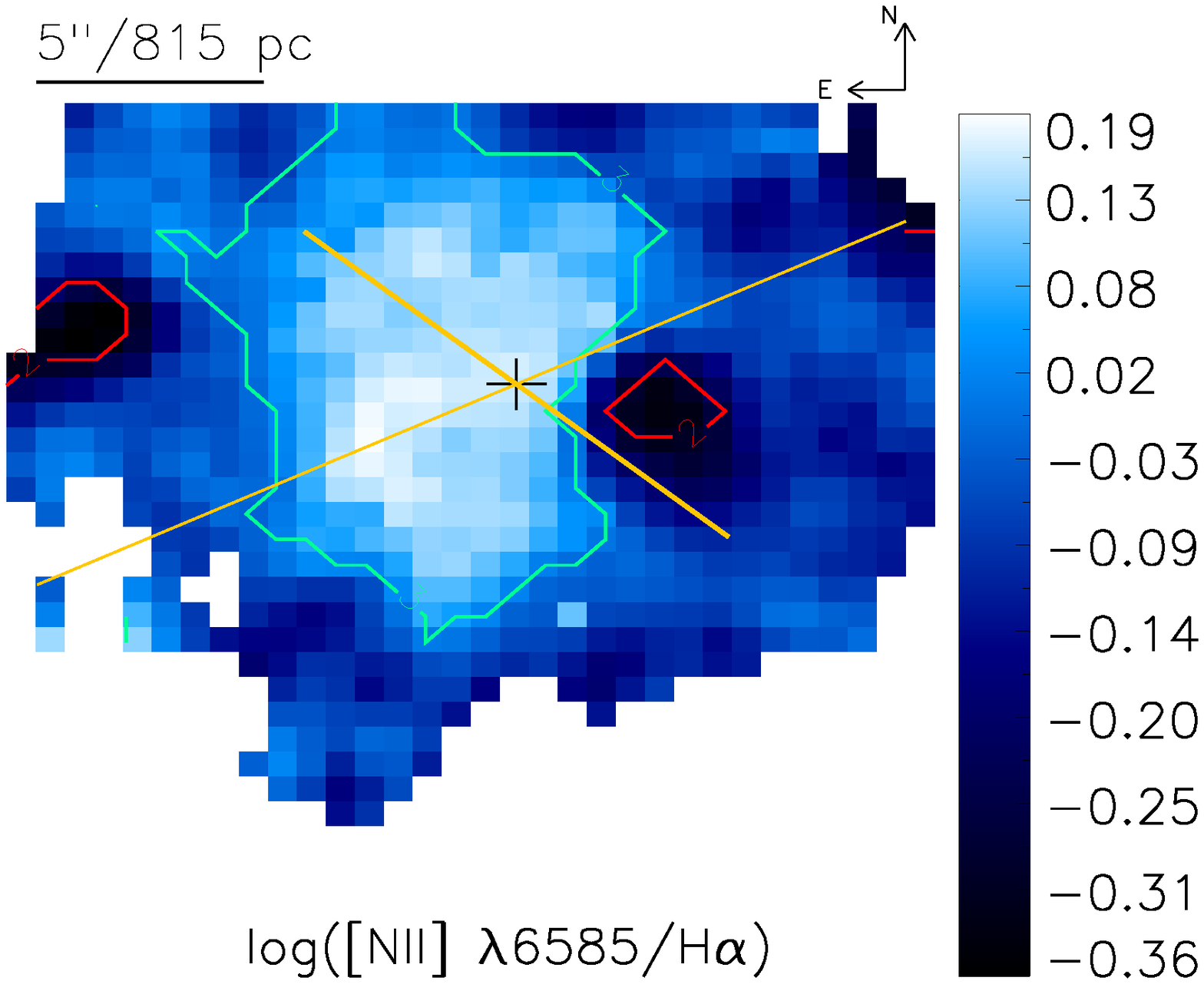}
  \includegraphics[width=0.32\textwidth]{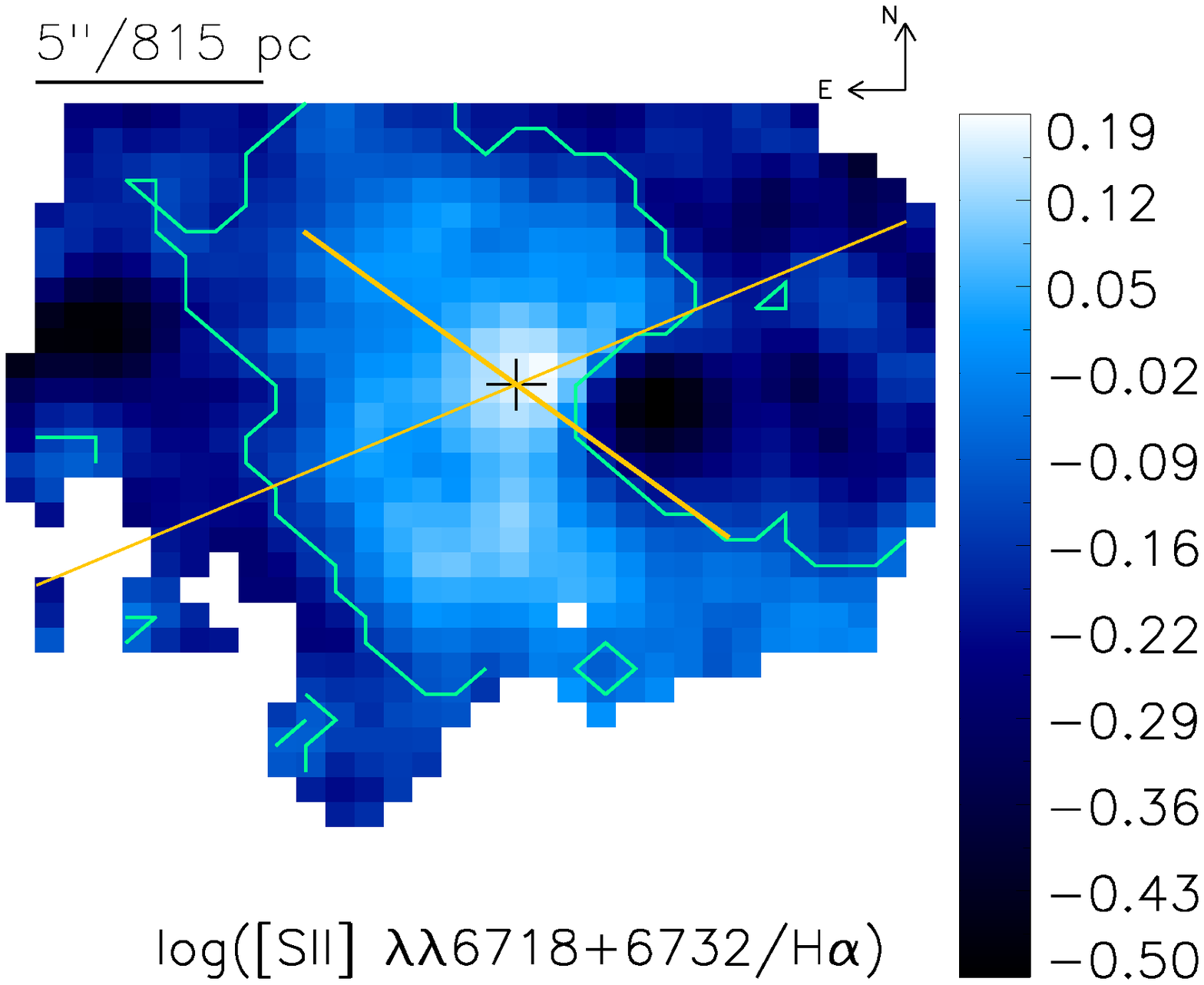}
  \caption{Maps of logarithmic emission line ratios. \textbf{Left:} The $\log\mathrm{\left([OIII]/H\beta\right)}$ map. Areas encircled with red contours are classed as SF-regions within the diagnostic diagrams. Areas enclosed by light green contour lines are classed as LINER-like. The remaining areas are diagnosed to be ionized of a composition of SF and AGN/LINER-like ionization mechanism. The classification made use of the demarcation lines by \cite{Kewley_NII, Kauffmann}. \textbf{Center:} The $\log\mathrm{\left([NII]/H\alpha\right)}$ map. The contours are the same as in the left panel. \textbf{Right:} The $\log\mathrm{\left([SII]/H\alpha\right)}$ map. The area enclosed in the contour line is classed as LINER-like according to the demarcation lines by \cite{Kewley_SII, Kauffmann}. The remaining parts are classed as SF.}
 \label{fig:diag_maps}
 \end{figure}

The ionized gas within the FOV is largely classed as `composite' and `LINER-like' in terms of diagnostic diagrams \cite{BPT,VO_diag} and demarcation lines within them \cite{Kewley_NII, Kewley_SII, Kauffmann}. The emission line ratio maps in Fig. \ref{fig:diag_maps} show the LINER-like emission to be extended predominantly along the inner bar. This becomes especially apparent in the $\log\mathrm{\left([SII]/H\alpha\right)}$ map. There the extension along the secondary bar is only interrupted by the presence of the central HII-region.\

Since the LINER-like emission is strongly extended, not only a photoionizing point source like an AGN is possible but extended/distributed ionization sources can be responsible as well. Commonly, this invokes hot evolved stellar populations (post-AGB) or shocks. They can ionize the gas in such a way that it mimics the emission pattern of LINERs. \cite{Cid_diag} found empirically that the LINER-like emission from retired galaxies, whose ISM is significantly ionized by p-AGB stars, is characterized by the equivalent width of H$\alpha<3$~\AA. This is given throughout the complete field of view (FOV) of our observations as well, with the central starforming-region (SF) as the only exception.\ 

Almost for sure shocks are always present in galaxies. However, the relatively flat LOSV dispersion map does not allow to estimate the contribution of slow shocks to the emission. As mentioned before, the regions of high dispersion values are coinciding with steep velocity gradients. Higher spatial resolution observations are necessary to assess their origin.\

Not only the LINER-like emission is extended. Interestingly, large areas of the FOV are classed as `composite'. Here we believe the line ratios to be caused by a mixture of the extended LINER-like emission with low-level SF.\

\begin{figure}
  \centering
  \includegraphics[width=0.32\textwidth]{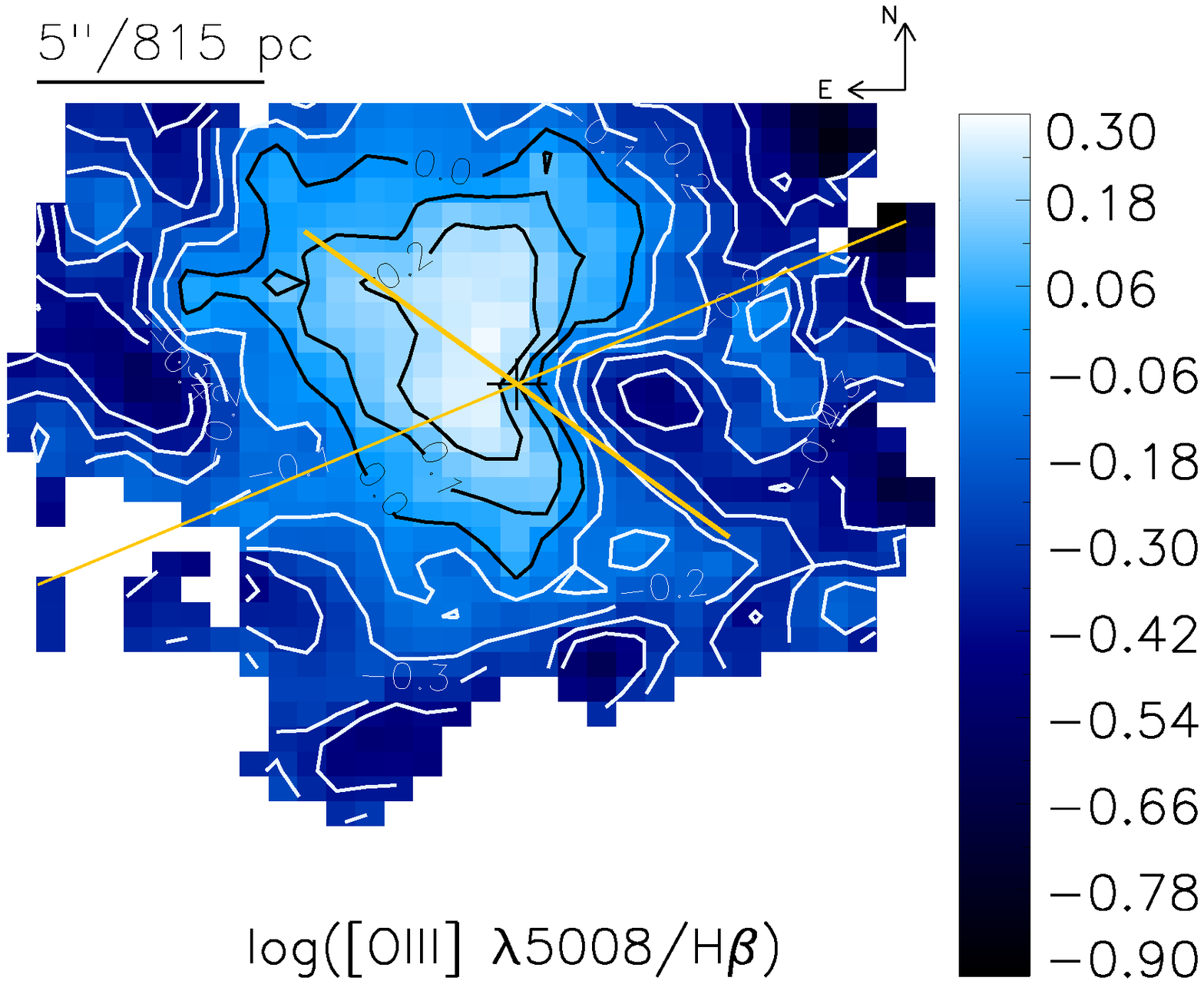}
  \includegraphics[width=0.30\textwidth]{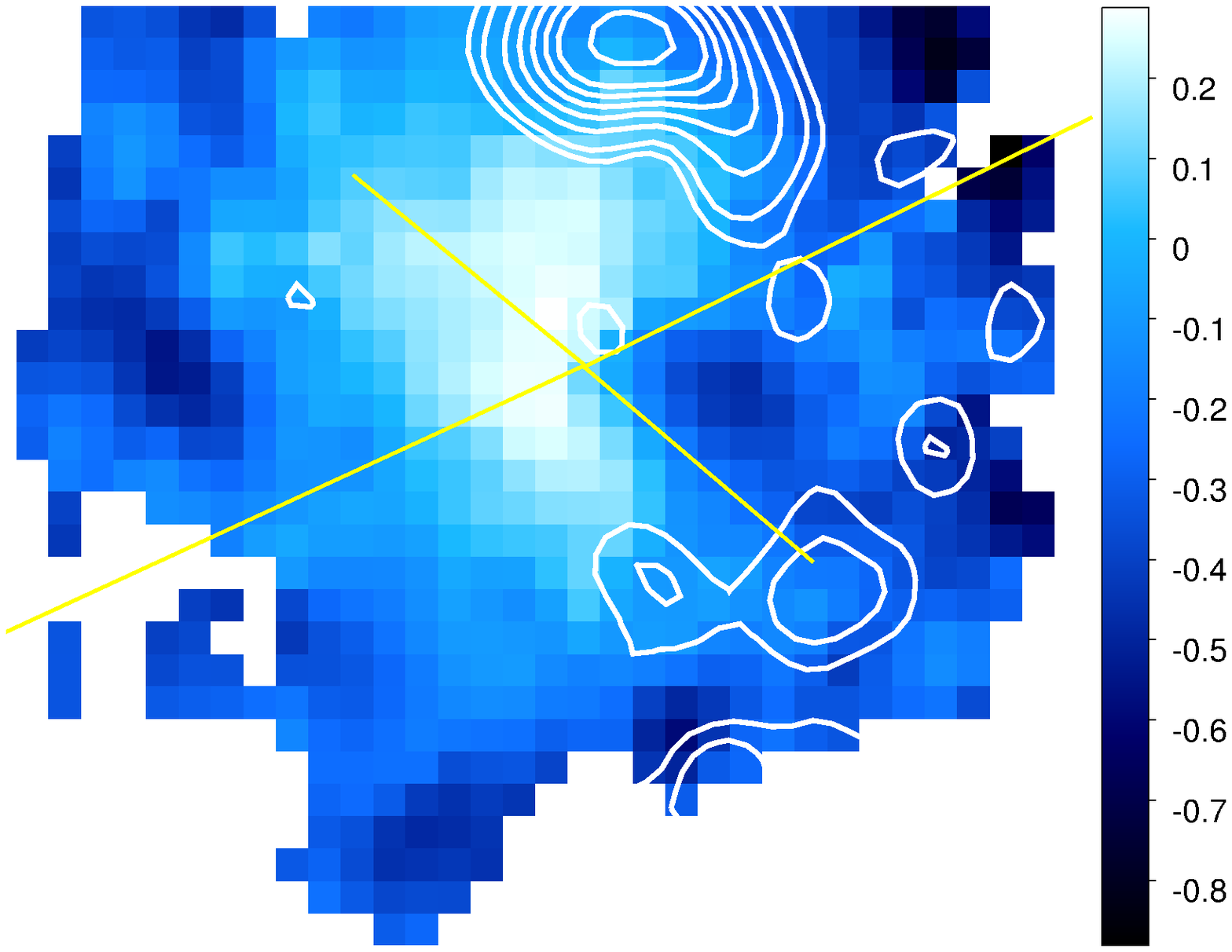}
  \caption{\textbf{Left:} The same as in left panel of Fig. 4 but with overplotted isocontour lines. \textbf{Right:} The same as in left panel but with overplotted $^{12}\mathrm{CO}$ radio intensity contours at 0.006, 0.01, 0.02, 0.03, 0.04, 0.05, 0.06, 0.08 and 0.1 Jy/beam from \cite{Leon}.}
 \label{fig:OIII_diag}
 \end{figure}

A closer look on the structure of the emission line ratio maps reveals a decrease of the line ratios radially outwards from the central SF-region. Exemplarily, we show the $\log\mathrm{\left([OIII]/H\beta\right)}$ map with overplotted isocontour lines in the left panel of Fig. \ref{fig:OIII_diag}. Possibly the central HII-region shields the radially more distant gas from the emission originating from the center. $^{12}\mathrm{CO}$ observations do not detect strong CO concentration in this area \cite{Leon} (see right panel of Fig. \ref{fig:OIII_diag}). Therefore, extinction effects seems not to play a significant role. The optical line ratios consist of emission lines spectrally close to each other and are consequently insensitive to reddening anyhow. However, mostly this feature vanishes within the uncertainties but it stays remarkable that it can be found in all three line ratio maps.

\section{Conclusions}
\label{conclusions}

The LOSV map of the gas shows a large kinematically decoupled core (KDC). The spiral structure within the KDC is strongly twisted. This might have its origin in the lopsidedness gas distribution on galactic scale \cite{Higdon} and within the KDC \cite{Leon}. In the center we find the s-shaped signature introduced by the gaseous secondary bar. The strong twist makes the inflow of gas towards the center a possible option and emphasizes the importance of secondary bars for AGN feeding.\

The LINER-like emitting gas is extended and ionized by distributed ionization sources, presumably p-AGB stars within the secondary bar. Slow shocks might play a role as well as the possible presence of a weak AGN. We ascribe the extended composite emission to a mixture of the mentioned LINER-like emission and photoionization by low-level star formation.\

{\scriptsize
 \bibliographystyle{abbrv}

}

\end{document}